\documentclass[runningheads]{llncs}
\usepackage[T1]{fontenc}
\usepackage{graphicx}
\usepackage{url}
\usepackage{listings}
\usepackage{booktabs}

\usepackage{xcolor}

\definecolor{codegreen}{rgb}{0,0.6,0}
\definecolor{codegray}{rgb}{0.5,0.5,0.5}
\definecolor{codepurple}{rgb}{0.58,0,0.82}
\definecolor{backcolour}{rgb}{0.95,0.95,0.92}

\lstdefinestyle{mystyle}{
    backgroundcolor=\color{backcolour},   
    commentstyle=\color{codegreen},
    keywordstyle=\color{magenta},
    numberstyle=\tiny\color{codegray},
    stringstyle=\color{codepurple},
    basicstyle=\ttfamily\footnotesize,
    breakatwhitespace=false,         
    breaklines=true,                 
    captionpos=b,                    
    keepspaces=true,                 
    numbers=left,                    
    numbersep=5pt,                  
    showspaces=false,                
    showstringspaces=false,
    showtabs=false,                  
    tabsize=2
}

\newcommand{\textcode}[1]{\texttt{\small{#1}}}

\begin{document}
\title{Streamlining Knowledge Graph Creation with PyRML}
%
%
\author{Andrea Giovanni Nuzzolese\inst{1}\orcidID{0000-0003-2928-9496}}
\authorrunning{A.G. Nuzzolese}
%
\institute{CNR - Institute of Cognitive Sciences and Technologies, Bologna, Italy\\
\email{andreagiovanni.nuzzolese@cnr.it}
}
\maketitle              
\begin{abstract}
Knowledge Graphs (KGs) are increasingly adopted as a foundational technology for integrating heterogeneous data in domains such as climate science, cultural heritage, and the life sciences. Declarative mapping languages like R2RML and RML have played a central role in enabling scalable and reusable KG construction, offering a transparent means of transforming structured and semi-structured data into RDF. In this paper, we present PyRML, a lightweight, Python-native library for building Knowledge Graphs through declarative mappings. PyRML supports core RML constructs and provides a programmable interface for authoring, executing, and testing mappings directly within Python environments. It integrates with popular data and semantic web libraries (e.g., Pandas and RDFlib), enabling transparent and modular workflows. By lowering the barrier to entry for KG creation and fostering reproducible, ontology-aligned data integration, PyRML bridges the gap between declarative semantics and practical KG engineering.
\end{abstract}
\section{Introduction}
\label{sec:intro}
Knowledge Graphs~\cite{Hogan_2022} (KGs) have become a foundational technology for integrating and querying heterogeneous data across domains such as climate science, cultural heritage, and life sciences. A core strength of KGs lies in their ability to make data explicit, interoperable, and semantically rich, aligning with the principles of FAIR~\cite{Wilkinson2016} (Findable, Accessible, Interoperable, and Reusable) data. 

Among the various approaches to constructing KGs, declarative mapping languages, such as R2RML\footnote{\url{http://www.w3.org/TR/r2rml/}} and RML~\cite{Dimou2014,Molina2023}, have emerged as key enablers in both literature and practice. By explicitly stating the rules for transforming data from structured and semi-structured sources (e.g., relational databases, CSV, JSON, XML, SPARQL services, etc.) into RDF, declarative mappings promote separation of concerns, reusability, and cross-source interoperability. These characteristics are particularly beneficial for collaborative and long-lived data integration efforts, where mapping logic must be shared, adapted, and audited over time. 

A variety of RML-compliant engines have been developed to support the execution of declarative mappings. Examples are RMLMapper\footnote{\url{https://github.com/RMLio/rmlmapper-java}}, CARML\footnote{\url{https://github.com/carml/carml}}, SDM-RDFizer~\cite{Iglesias2025}, and Morph-KGC~\cite{Arenas2024}. These tools have proven effective in translating structured data into RDF at scale and have contributed significantly to the maturation of the semantic data integration ecosystem. However, their usage often assumes familiarity with command-line interfaces, specific configuration formats, or specific programming language environments, which can present barriers to integration in modern data science workflows. In addition, features such as mapping modularity, incremental generation, unit testing, and tight coupling with ontological reasoning are still underdeveloped or inconsistently supported across tools. In this context, PyRML is conceived as a Python-native alternative that supports interactive, programmable, and transparent KG construction. It complements existing engines while focusing on usability, extensibility, and seamless integration with the Python data ecosystem. Additionally, by abstracting some of the technical complexity while maintaining expressive power, PyRML contributes to bridging the gap between declarative semantics and practical KG engineering.

The remainder of this paper is the following. Section~\ref{sec:soa} provides an overview of the related work. Section~\ref{sec:system} details the proposed system architecture with usage examples. Section~\ref{sec:eval} describes the evaluation methodology and results. Finally, Section~\ref{sec:conclusion} discusses conclusions and future directions.

\section{Related work}
\label{sec:soa}
The construction of KGs from structured and semi-structured data has been extensively explored in the Semantic Web community\footnote{Please refer to \url{https://kg-construct.github.io/awesome-kgc-tools/} for an overview on existing tools for KG construction.}. 
The foundational approaches focused on the integration of semantic data. These include the use of a view-based paradigm~\cite{Lenzerini2002}, such as, Global-As-View~\cite{Halevy2001} (GAV), Local-As-View~\cite{Ullman2000} (LAV), and Global-Local-As-View~\cite{Friedman1999} (GLAV). A view defines the relationships between heterogeneous data sources and a unified mediated schema. These paradigms, originally developed in the context of data warehouse and federated databases, provided the theoretical foundation for later declarative mapping languages used in the construction of KG. In particular, the GAV approach, where each element of the mediated schema is defined as a query over the sources, closely resembles modern RML mappings, where ontology terms are defined in terms of data source structure. In contrast, LAV and GLAV underpin more expressive approaches such as Ontology-Based Data Access (OBDA), where mappings specify how source data can satisfy arbitrary ontology queries. Although powerful, OBDA systems often rely on complex reasoning services, which can hinder scalability and accessibility for practitioners~\cite{Lembo2015}.

In recent years, declarative mapping languages such as the RDB to RDF Mapping Language (R2RML) and RDF Mapping Language~\cite{Dimou2014,Molina2023} (RML) have become a standard mechanism for aligning raw data with RDF vocabularies and ontologies in a transparent and maintainable way. More specifically R2RML is designed to cope with the transformation of relational databases to RDF, whilst RML generalises the mapping model of R2RML to support diverse semi-structured data sources.
Accordingly, several tools have been developed to implement and execute RML mappings. The RMLMapper, written in Java, was among the first engines to support full RML core semantics and has been widely used for research and data publication tasks. CARML builds on the same paradigm, offering improved modularity and performance. More recently, tools like SDM-RDFizer~\cite{Iglesias2025} and Morph-KGC~\cite{Arenas2024} have focused on scalability and performance, enabling the efficient generation of large KGs from relational databases and tabular data. These engines have been successfully applied in large-scale projects, such as iASiS\footnote{\url{http://project-iasis.eu/}}, which is an EU funded project to enable precision medicine approaches by utilising insights from patient data.

Despite their robustness, existing RML engines often assume specific technological stacks (e.g., Java or Docker-based deployments), and their integration with modern data science workflows—typically centred around Python—is limited. PyRML contributes to this landscape by offering a Python-native, programmable interface for RML-based KG construction. Unlike black-box engines, PyRML enables fine-grained control over mapping composition, execution, and testing, and is designed to integrate with widely used Python libraries such as Pandas and RDFlib. This positions PyRML as a complementary tool in the RML ecosystem, addressing the need for flexible, scriptable, and developer-friendly solutions in data-centric environments.

A complementary approach to data integration is presented by SPARQL Anything~\cite{Asprino2023}, which enables querying heterogeneous data sources directly using SPARQL, without the need for upfront data transformation into RDF. By overloading the SERVICE clause in SPARQL 1.1, SPARQL Anything allows users to access data from various formats through a uniform SPARQL interface. This approach leverages the Facade-X meta-model~\cite{Daga2021} to provide a simplified RDF representation of diverse data sources, facilitating rapid prototyping and ad-hoc querying. While SPARQL Anything excels in on-the-fly data access, it does not produce persistent RDF graphs, which may be a limitation for applications requiring long-term data storage and reasoning capabilities.

\section{The PyRML system}
\label{sec:system}
\subsection{Architecture}
\begin{figure}
    \centering
    \includegraphics[width=1\linewidth]{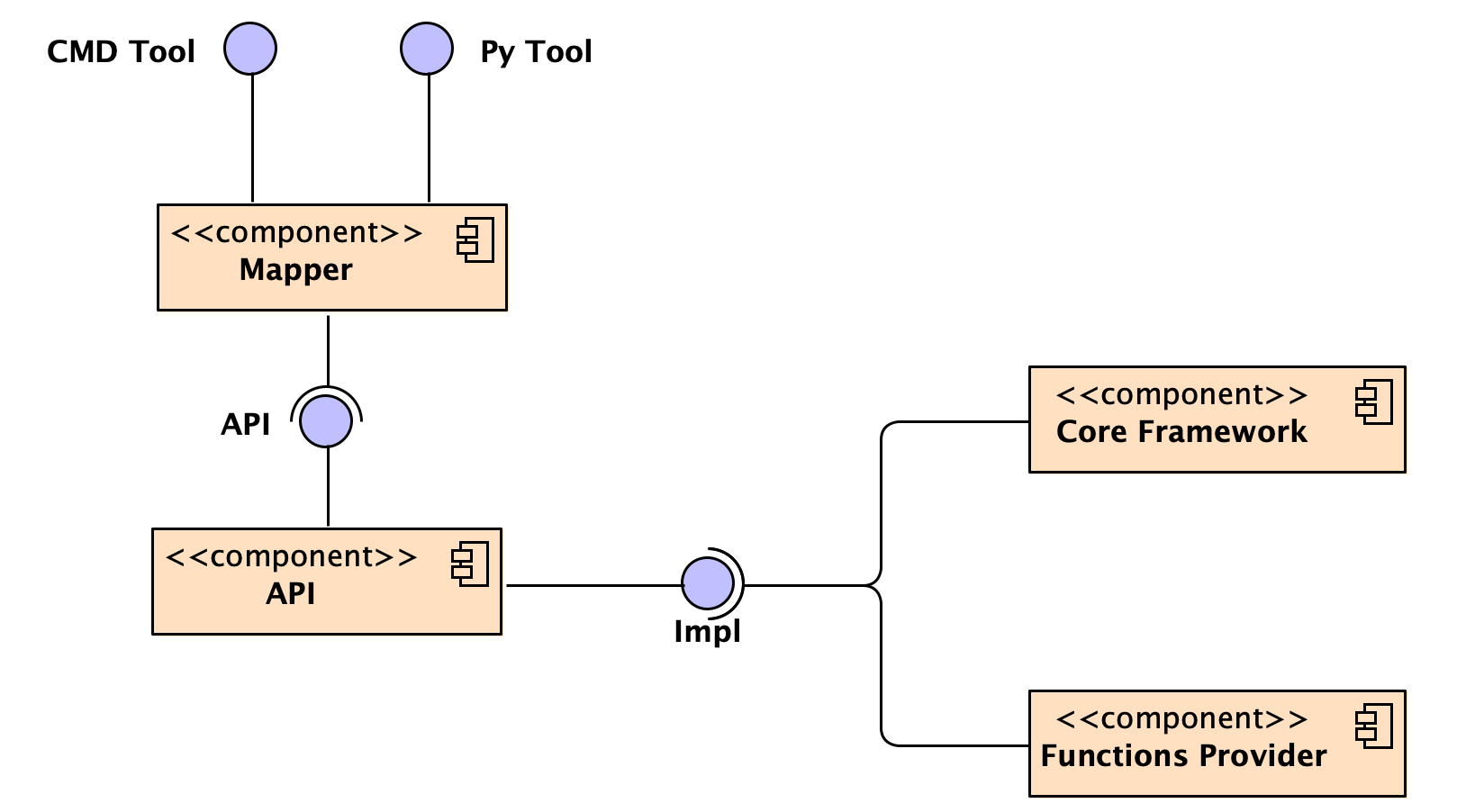}
    \caption{The architecture of PyRML}
    \label{fig:architecture}
\end{figure}

Figure~\ref{fig:architecture} shows the modular architecture of PyRML that counts of four main modules. These modules are: (i) the API module, (ii) core Framework, (iii) Functions Provider, and (iv) Mapper. 

\paragraph{\bf API module.} The API module provides the abstract base classes that define the core structure of the programming interface for capturing the RML model within the software platform. At the top of this structure is the \textcode{TermMap} abstract base class, which represents any entity in an RML mapping associated with an IRI and intended for generating RDF data from a logical table. It is worth noting that the class hierarchy defined in PyRML directly mirrors the taxonomy of classes specified in the RML and R2RML ontologies. Consequently, examples of classes derived from \textcode{TermMap} include \textcode{SubjectMap}, which specifies the mapping instructions to generate the subject of a triple from a logical table, and \textcode{PredicateObjectMap}, which links a \textcode{PredicateMap} and an \textcode{ObjectMap} to generate the predicate and object of a triple, respectively.

The \textcode{TermMap} class extends the abstract base class \textcode{IdentifiedNode}, implemented by RDFLib\footnote{\url{https://github.com/RDFLib/rdflib}}, a widely used Python package for working with RDF. Hence, all instances of \textcode{TermMap} are valid RDF terms that can be used as the subject, predicate, or object of an RDF triple when constructing an RDFLib graph.

For example, the following code defines a \textcode{TermMap} through its derived class \textcode{SubjectMap}, and uses it as the subject of a triple to construct a graph with RDFLib.
\vspace{0.4cm}
{
\begin{lstlisting}[language=Python]
from pyrml import TermMap, SubjectMap
from rdflib import Graph, Namespace, RDF

ex: Namespace = Namespace('https://foo.org/example/')
rr: Namespace = Namespace('http://www.w3.org/ns/r2rml#')
tm: TermMap = SubjectMap(ex.SM)

g: Graph = Graph()
g.add((tm, RDF.type, rr.SubjectMap))
\end{lstlisting}
}
\vspace{0.4cm}
The abstract methods of TermMap include: (i) \textcode{to\_rdf}, which converts the PyRML term into an RDFLib graph while preserving the graph structure rooted at that term; (ii) \textcode{apply}, which performs the mapping by using all the informations associated with a term against a given \textcode{LogicalSource}; and (iii) \textcode{from\_rdf}, which is a static method that allows to instantiate a \textcode{TermMap} and its associated terms directly from an RDFLib graph. These three abstract methods are implemented by the base classes of \textcode{TermMap} that are defined in the core molude.

\paragraph{\bf Core module.} The core module has a twofold purpose, as reflected in its name: (i) it serves as the foundation of PyRML by implementing all the derived classes of \textcode{TermMap} that enable PyRML’s functionality, and (ii) it addresses the core model of RML.
The constructor (i.e. the \textcode{\_\_init\_\_} method) of the class \textcode{TermMap} accepts a mandatory positional argument, which is the IRI of an RML term, and optional keyword arguments that can be used to associate term-specific values with an RML term, such as the values of the \textcode{rr:template} or \textcode{rml:reference} predicates in an RML mapping.
The following code snippet defines RML triples map in a programmatic way.
\vspace{0.5cm}
{
\begin{lstlisting}[language=Python]
from rdflib import FOAF

ls: LogicalSource = ... // details on logical sources later

sm: SubjectMap = SubjectMap(ex.SM,
                         template='https://foo.org/d/{ID}',
                         _classes=FOAF.Person)

pm: PredicateMap = PredicateMap(ex.PM, constant=FOAF.name)

om: ObjectMap = ObjectMap(ex.OM,
                          reference='name',
                          term_type=rr.Literal)

pom: PredicateObjectMap = PredicateObjectMap(ex.POM, 
                                             predicates=pm,
                                             object_maps=om)

tm: TripleMappings = TripleMappings(ex.Tm,
                                logical_sources=ls,
                                subject_maps_mp,
                                predicate_object_maps=pom)

g: Graph = tm.to_rdf()
\end{lstlisting}
}
\vspace{0.4cm}
In the code snippet above a \textcode{SubjectMap} named \textcode{ex:SM} is created at line 5. This \textcode{SubjectMap} uses the template IRI \textcode{https://foo.org/d/\{ID\}}, where \textcode{\{ID\}} is replaced by the value of the ID from the input data. It also declares that each generated subject is an instance of \textcode{foaf:Person}, specified via the RML declaration \textcode{\_class=FOAF.Person}. 

Instead, an instance of \textcode{PredicateMap} is defined at line 9. This indicates that the predicate of the triple is the FOAF property \textcode{foaf:name}. Then, an \textcode{ObjectMap} named \textcode{ex:OM} is instantiated at line 11. The object value is taken from the name attribute in the input data (i.e. \textcode{reference='name'}) and the term type is set to be an RDF literal (i.e. \textcode{term\_type=rr.Literal}). At line 15 a \textcode{PredicateObjectMap} is created by connecting the \textcode{pm} predicate to the \textcode{om} object map. Finally, a triples mapping is instantiated at line 19. The latter represents a complete mapping rule that takes data from the logical source identified by \textcode{ls}, builds a subject from \textcode{sm}, and predicates and objects from \textcode{pom}.
An invocation of the method \textcode{to\_rdf()} against the triples mapping \textcode{tm} is provided at line 24 to return an RDFLib graph as output. This graph is represented in the Turtle serialisation below.
\vspace{0.4cm}
{
\begin{lstlisting}
@prefix ex: <https://foo.org/example/> .
@prefix rr: <http://www.w3.org/ns/r2rml#> .
@prefix rml: <http://semweb.mmlab.be/ns/rml#> .
@prefix foaf: <http://xmlns.com/foaf/0.1/> .

ex:Tm a rr:TriplesMapping ;
      rml:logicalSource ... ;
      rr:subjectMap ex:SM ;
      rr:predicateObjectMap ex:POM .

ex:SM a rr:SubjectMap ;
      rr:template 'https://foo.org/d/{ID}' ;
      rr:class foaf:Person .

ex:PM a rr:PredicateMap ;
      rr:constant foaf:name .
      
ex:OM a rr:ObjectMap ;
      rml:reference 'name' ;
      rr:termType rr:Literal .

ex:POM a rr:PredicateObjectMap ;
      rr:predicateMap ex:PM ;
      rr:objectMap ex:OM .
\end{lstlisting}
}
\vspace{0.4cm}
As stated previously, any instance of \textcode{TermMap} can be generated directly from an RDFLib graph by invoking the \textcode{from\_rdf()} method on the corresponding class. For example, the triples map \textcode{ex:TM} can be converted to a PyRML object as shown in the code block below.
\vspace{0.4cm}
{
\begin{lstlisting}[language=Python]
tm: TripleMappings = TripleMappings.from_rdf(g, parent=ex.TM)
\end{lstlisting}
}
\vspace{0.4cm}
The execution of the declarative mappings is enabled by the method \textcode{apply} that performs tranformation defined in a \textcode{TermMap} against a specific \textcode{LogicalSource}. In PyRML, a \textcode{LogicalSource} makes use of a Pandas DataFrame for providing a flexible and Python-native abstraction of the input data. This design decouples data access and parsing from the mapping logic, allowing users to load, clean, transform, and prepare data using familiar Pandas operations before applying declarative mappings. By working with DataFrames, PyRML integrates seamlessly into Python-based data science workflows, enabling direct manipulation of data, efficient debugging, and reuse of in-memory datasets from diverse sources (e.g., CSV, databases). This approach aligns the RML notion of logical tables with the DataFrame’s tabular structure, facilitating transparent and programmable knowledge graph construction while maintaining compatibility with RML’s mapping semantics. In this context, the \textcode{apply} method enables the vectorised application of transformation operations across a DataFrame that represents a \textcode{LogicalSource}. Instead of processing records one by one, the \textcode{apply} method efficiently computes the corresponding RDF terms (e.g., subject IRIs, object literals, etc.) for all rows in a single operation, leveraging Pandas' inherent performance optimisations. This vectorised processing enhances scalability and supports the integration of complex transformation logic directly into the mapping execution pipeline.

PyRML currently supports the following data sources for instantiating a \textcode{Logical\-Source}: (i) CSV, (ii) XML, (iii) JSON, (iv) SPARQL, (v) MySQL, (vi) SQL Server, and (vii) PostgreSQL. A \textcode{Logical\-Source} in PyRML is built on top of a \textcode{Source}, which is the base abstract class used to represent various data source types. For instance, the class \textcode{CSV\-Source} extends \textcode{Source} to model a CSV data source. The code snippet below illustrates how to create a logical source from a CSV file.
\vspace{0.4cm}
{
\begin{lstlisting}[language=Python]
from pyrml import LogicalSource, Source, CSVSource

source: Source = CSVSource(ex.CSV, 'students.csv')
ls: LogicalSource(ex.LS, sources=source)
\end{lstlisting}
}
\vspace{0.4cm}
\paragraph{\bf Functions module.} Many real-world scenarios require additional data transformations beyond simple attribute retrieval or template substitution---such as string manipulation, date formatting, or value normalisation. To address these needs, the RML community introduced RML Functions\footnote{\url{https://kg-construct.github.io/rml-fnml/spec/docs/}} (or FnO Functions), a mechanism to declaratively invoke functions as part of a mapping. PyRML provides a functions module to support the use of RML Functions---operations that enable data transformations within declarative mappings, following the principles of the Function Ontology (FnO). Similar to engines like RMLMapper, PyRML includes a list of built-in functions that mirror the default set provided by RMLMapper, allowing users to leverage standard transformation operations out of the box. This list is implemented in the Function module\footnote{\url{https://github.com/anuzzolese/pyrml/blob/master/pyrml/functions.py}}.

A distinctive feature of PyRML is its support for user-defined functions. PyRML allows developers to extend the set of available functions programmatically, integrating custom transformation logic directly into the mapping execution pipeline. This is achieved by defining a standard Python function and registering it using a dedicated decorator provided by the library, called \textcode{rml\_function}.
The \textcode{rml\_function} decorator facilitates the registration of a Python function as an RML Function by associating it with a function identifier (IRI) and parameter mappings. Its implementation follows a standard Python decorator pattern, wrapping the original function and registering it with the PyRML runtime. Once registered, the function can be invoked within RML mappings through the Function Ontology (FnO) mechanism. The following are the signature and a usage example of the \textcode{rml\_function} decorator.
\vspace{0.4cm}
{
\begin{lstlisting}[language=Python]
def rml_function(fun_id: str, **params: Dict[str, str]) -> Callable:
    ...

@rml_function(
 fun_id='http://users.ugent.be/~bjdmeest/function/grel.ttl#toLowerCase',
 value='http://users.ugent.be/~bjdmeest/function/grel.ttl#valueParameter')
def to_lower_case(value: str) -> str:
    return value.lower()
\end{lstlisting}
}
\vspace{0.4cm}
In the example above the Python function \textcode{to\_lower\_case} implements the logic to convert a string to lowercase. The \textcode{@rml\_function} decorator registers it under the IRI corresponding to the GREL\footnote{\url{https://openrefine.org/docs/manual/grelfunctions}} \textcode{toLowerCase} function from FnO. The mapping between the FnO parameter IRI and the Python argument name is specified via the value keyword. Once registered, this function can be invoked inside an RML mapping referencing its function IRI, enabling seamless integration between declarative mappings and Python-defined transformation logic.

\paragraph{\bf Mapper module.} The mapper module is the execution core of PyRML, responsible for transforming RML mapping definitions into RDF triples. It provides a Python-native, extensible, and efficient engine for materialising knowledge graphs from structured and semi-structured data, fully aligned with the RDF Mapping Language (RML). This module enables the seamless integration of declarative knowledge graph construction into Python-based workflows.
At its core is the \textcode{RMLConverter} class, which implements the high-level interface for executing a set of RML mappings. It supports both single-threaded and parallel execution strategies, allowing it to scale from lightweight testing to larger data transformation tasks. When processing a mapping, the converter parses the mapping document, builds an internal representation of the mapping using the classes for term maps defined in the API and Core modules, and applies each term map to the associated logical source by leveraing Pandas. When processing a mapping, the converter first renders the RML document using Jinja2\footnote{https://jinja.palletsprojects.com/en/stable/} templating, which allows users to define parameterised and reusable mappings. Template variables can be injected at runtime, enabling dynamic substitution of values such as file paths, graph names, or filter conditions, improving the flexibility and maintainability of mapping definitions. The class \textcode{RMLConverter} can be instantiated programmatically or used from command line through a dedicated Python script. The following is an example of the programmatic use of the class \textcode{RMLConverter}.
\vspace{0.4cm}
\begin{lstlisting}[language=Python]
from pyrml import PyRML, RMLConverter

c: RMLConverter = PyRML.get_mapper()

'''
Invoke the method convert on the instance of class RMLConverter by:
 - using the persons.ttl RML descriptor;
 - obtaining an RDF graph as output.
'''
g : Graph  = c.convert('persons.ttl')
\end{lstlisting}
\vspace{0.4cm}
Variables for Jinja2 templating can be provided through the \textcode{template\-\_vars} argument of the \textcode{convert} method. This argument accepts a Python dictionary, where each key corresponds to a parameter referenced in the template, and each value specifies the actual value to substitute. 
An example of an RML file that makes use of templating is the following. 
\vspace{0.4cm}
\begin{lstlisting}
<#Mapping> a rr:TriplesMap;
  rml:logicalSource [
    rml:source "{{ INPUT_CSV }}" ;
    rml:referenceFormulation ql:CSV
  ];
  rr:subjectMap [
    ...
  ] .
\end{lstlisting}
\vspace{0.4cm}
The template variable in the RML above is reported at line 3, i.e. \textcode{\{\{ INPUT\_CSV \}\}} and can be set to an actual value as in the following example.
\vspace{0.4cm}
\begin{lstlisting}[language=Python]
vars = {'INPUT_CSV': 'students.csv'}
g : Graph  = c.convert('persons.ttl', template_vars=vars)
\end{lstlisting}
\vspace{0.4cm}
We note that Jinja2 templating, as implemented in PyRML, is a non-standard extension and is not part of the official RML specification or reference documentation. For more details on how templating works, please refer to the official Jinja2 documentation.

Instead, the following is a usage example of the command line tool \textcode{pyrml-mapper.py} that wraps the application aroung the \textcode{RMLConverter}. 

\vspace{0.4cm}
\begin{lstlisting}[language=Bash]
python pyrml-mapper.py [-o RDF out file] [-f RDF out file] [-m] input
\end{lstlisting}
\vspace{0.4cm}
Where:
\begin{itemize}
  \item \textcode{input} is the required positional argument that specifies the RML mapping file to be used for RDF conversion;
  \item \textcode{-o filename} is an optional argument that specifies the output file for saving the resulting RDF graph. If omitted, the output is written to standard output by default;
  \item \textcode{-f rdf-syntax} is an optional argument to define the syntax used to serialise the RDF graph. Supported values include n3, nquads, nt, pretty-xml, trig, trix, turtle, and xml. If not specified, nt (N-Triples) is used by default;
  \item \textcode{-m} is an optional flag that enables multiprocessing to accelerate the transformation process.
\end{itemize}

\subsection{Release and Availability Notes}
\label{sec:release-notes}
\paragraph{\bf Reusability.} PyRML is released as open-source software under the Apache 2.0 license\footnote{\url{https://github.com/anuzzolese/pyrml?tab=Apache-2.0-1-ov-file}}. It is well-documented, with API references, examples, and tutorials available in the GitHub repository. It is general-purpose and not tied to any specific domain, making it suitable for a wide range of use cases. The modular architecture supports extension, e.g., custom functions and mapping loaders. The documentation clearly specifies supported features, usage patterns, and known limitations, allowing users to adapt the tool confidently to their own needs.
\paragraph{\bf Availability.}
PyRML is publicly available at:
\begin{itemize}
    \item GitHub: \url{https://github.com/anuzzolese/pyrml};
    \item Python Package Index (PyPI): \url{https://pypi.org/project/pyrml-lib}, which allows to download and install PyRML directly from the official third-party software repository for Python via the \textcode{pip} command, e.g. \textcode{pip install pyrml-lib};
    \item Documentation: included in the repository and browsable via 
    GitHub Pages;
    \item Canonical citation: DOI: \url{https://doi.org/10.5281/zenodo.15399948} released by Zenodo;
    \item License: Apache 2.0.
\end{itemize}

\paragraph{\bf Impact and adoption.} PyRML fills a critical gap in the knowledge graph construction landscape by providing a Python-native, declarative RML mapping engine that integrates seamlessly with modern data science workflows. It is of direct relevance to the Semantic Web community, particularly to researchers and practitioners engaged in FAIR data, open science, and knowledge integration initiatives. More broadly, PyRML lowers the barrier to entry for the scientific and societal adoption of Semantic Web technologies by aligning with widely adopted tools and conventions in the Python ecosystem.
PyRML has already seen adoption in several EU-funded projects. It has been successfully integrated into the data transformation pipelines of HACID (Hybrid Human-AI Collective Intelligence in Open-Ended Domains), WHOW (Water Health Open Knowledge), and FOSSR (Fostering Open Science in Social Sciences and Humanities). In these projects, PyRML has been used to support transparent, modular, and reproducible workflows for transforming heterogeneous data sources into semantically rich RDF graphs aligned with domain ontologies. Its programmable and extensible architecture has proven particularly valuable in collaborative and evolving research environments.
The project is actively maintained by a team of four developers and is openly available on GitHub. As of May 13th, 2025, the repository has received 37 stars, been forked 13 times, and, between April 30th and May 13th, has been cloned 30 times and viewed 155 times.\footnote{Usage statistics are available at \url{https://github.com/anuzzolese/pyrml/graphs/traffic}.}
These indicators of adoption and engagement confirm the practical relevance of PyRML and its growing role in enabling declarative, transparent, and reproducible knowledge graph engineering across both research and applied domains.

\section{Evaluation}
\label{sec:eval}
\subsection{Experimental setup}
\label{sec:eval-experimental-setup}
To evaluate the correctness and performance of PyRML, we designed an experimental protocol based on the official RML-Core test cases\footnote{\url{https://github.com/RMLio/rml-test-cases}}. The RML-Core test cases are a standardised suite developed by the Knowledge Graph Construction Community Group\footnote{\url{https://www.w3.org/community/kg-construct/}} to evaluate the correctness and feature coverage of RML-compliant engines. Each test case defines a mapping scenario with an input data source, an RML mapping file, and an expected RDF output. The suite covers key RML features such as logical sources, templates, joins, constant values, and graph maps. These tests are designed to be minimal, deterministic, and interpretable, making them ideal for verifying whether an engine behaves according to the standard. Successful execution of the core test cases provides strong evidence of RML compliance and correctness.
Supported input formats in the test suite include CSV, JSON, XML, SPARQL endpoints, MySQL, SQL Server, and PostgreSQL databases. By validating against these tests, an engine demonstrates conformance to the RML specification across a wide range of source types and mapping patterns.The evaluation was carried out in two phases: feature coverage and computational performance benchmarking.

\paragraph{\bf RML-Core conformance.} In the first phase, we assessed the coverage of PyRML against the RML-Core specification\footnote{\url{https://kg-construct.github.io/rml-core/spec/docs/}}. This was done by systematically executing all the RML-Core test cases and verifying whether the output RDF graphs conformed to the expected results. This step ensured that PyRML adheres to the semantics and structural requirements of the RML specification. For this analysis we set up a Docker container to provide Apache Jena Fuseki\footnote{\url{https://jena.apache.org/documentation/fuseki2/}} as SPARQL endpoint, MySQL, PostgreSQL, and SQL Server. The container is available on GitHub\footnote{\url{https://github.com/anuzzolese/pyrml-testing/tree/main/docker-framework}} and can be instantiated via \textcode{docker-compose}\footnote{Please refer to the readme provided along with the Docker container.} Then, a Python script\footnote{\url{https://github.com/anuzzolese/pyrml-testing/blob/main/unittesting.py}} was developed to assess RML-Core conformance using Python’s unit testing framework, where each RML-Core test case is interpreted as a separate unit test.

\paragraph{\bf Computational performance.} In the second phase, we benchmarked the computational performance of PyRML and compared it to the widely used RMLMapper engine. For each test case, we executed the transformation 10 times using both engines and recorded the execution time in milliseconds. The final reported time for each engine and test case corresponds to the average execution time over the 10 runs. This procedure reduces the impact of transient system-level fluctuations and provides a more stable basis for comparison. The bash script that enabled the comparison is available on GitHub\footnote{\url{https://github.com/anuzzolese/pyrml-testing/blob/main/comparative-analysis.sh}}.

All benchmarks were run under identical conditions on the same hardware environment to ensure comparability, that is a 2.3 GHz Quad-Core Intel Core i7 with 32 GB of memory. The results of this evaluation are reported in Section~\ref{sec:results}.

\subsection{Results}
\label{sec:results}
\paragraph{\bf RML-Core conformance.} Table~\ref{tb:core-coverage} presents the coverage of PyRML against the reference RML core test suite, broken down by the type of logical data source. Each row corresponds to a different source type supported in RML, and the table reports three key metrics: (i) the number of test cases successfully executed by PyRML, where the generated RDF output matches the expected result; (ii) the number of test cases where PyRML did not produce the expected output, either due to missing feature support or incorrect behaviour; and (iii) the total number of test cases defined in the suite for that source type.
\begin{table}[]
\centering
\caption{PyRML coverage of test cases for RML core.}
\label{tb:core-coverage}
\resizebox{.8\textwidth}{!}{
\begin{tabular}{p{3cm}p{2.5cm}p{2.5cm}p{2.5cm}}
\toprule
\textbf{Source type} & \textbf{Passed} & \textbf{Failed} & \textbf{\# of test cases}\\ 
\midrule
CSV & 39   & 0  & 39 \\
JSON & 40   & 0  & 40 \\
XML & 38   & 0  & 38 \\
SPARQL & 24   & 2  & 26 \\
MySQL & 55 & 5 & 60 \\
PostSQL & 55 & 5 & 60 \\
SQL Server & 55 & 5 & 60 \\
\bottomrule
\end{tabular}
}
\end{table}

Complete coverage (100\% pass rate) is achieved for CSV, JSON, and XML, indicating robust and stable support for these commonly used structured and semi-structured formats. Among the 26 RML core test cases involving SPARQL data sources, PyRML successfully passes 24 and fails 2. The two failed cases, i.e. \textcode{RMLTC0008b-SPARQL} and \textcode{RMLTC0009a-SPARQL}, show limitations in the current implementation of join semantics when SPARQL is used as a logical source. In fact, \textcode{RMLTC0008b-SPARQL} tests the generation of triples that involve a referencing object map, where data from one source must be joined with another in a specific predicate object map. Similarly, \textcode{RMLTC0009a-SPARQL} evaluates the handling of foreign key-style relations between logical sources---an operation conceptually analogous to joins in relational databases. While PyRML supports referencing object maps for sources such as CSV and JSON, support for such joins across SPARQL requires more investigation and testing. 
Among the 60 RML core test cases available for each of the relational database systems, i.e. MySQL, PostgreSQL, and SQL Server, PyRML passes 55 and fails 5, which are: (i) \textcode{RMLTC0009d}; (ii) \textcode{RMLTC0011a}; (iii) \textcode{RMLTC0013a}; (iv) \textcode{RMLTC0015a}; and (v) \textcode{RMLTC0016d}. These failures are consistent across all three systems and are attributed to specific limitations in the current implementation of PyRML's SQL mapping layer.
The test case \textcode{RMLTC0009d} checks the ability to handle column names that match SQL reserved keywords. PyRML currently does not implement automatic quoting or escaping of such identifiers, leading to parsing or execution errors during mapping. \textcode{RMLTC0011a} involves the mapping of many-to-many (M:N) relationships via custom SQL queries embedded in the logical source. \textcode{RMLTC0013a} tests the behaviour of referencing object maps when joined columns contain null values. In accordance with the RML specification, no triples should be generated in such cases; however, PyRML does not yet suppress triple generation in the presence of nulls, resulting in incorrect output. \textcode{RMLTC0015a} evaluates the correct generation of language-tagged literals based on values from a source column. Finally, \textcode{RMLTC0016d} tests the handling of datatype conversions, specifically boolean values. PyRML does not yet support automatic casting of source values to \textcode{xsd:boolean}, which leads to failures in producing semantically correct RDF literals when boolean datatypes are required.
All these test cases represent clear, bounded limitations rather than architectural constraints. All identified issues are part of PyRML’s ongoing development roadmap, and their resolution is planned in upcoming releases to support full RML compliance across relational database backends.

\begin{figure}
    \centering
    \includegraphics[width=.9\linewidth]{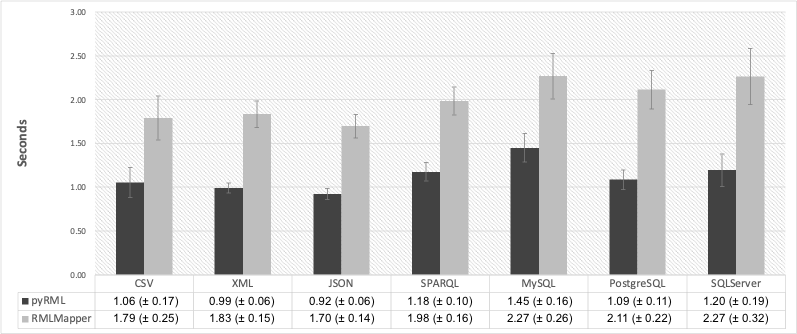}
    \caption{Comparison of PyRML with RMLMapper respect to execution time of test cases expressed in seconds.}
    \label{fig:eval-exec-times}
\end{figure}

\paragraph{\bf Computational performance.} Figure~\ref{fig:eval-exec-times} shows the results of the comparative analysis between PyRML and RMLMapper. The results are expressed in seconds, whilst error bars represent standard deviotions, which is reported among brackets.

The results demonstrate a consistent performance advantage for PyRML across all source types. For example, on CSV sources, PyRML achieved an average execution time of 1.06 seconds, compared to 1.79 seconds for RMLMapper. Similar gains are observed for XML (0.99s vs. 1.83s), JSON (0.92s vs. 1.70s), and SPARQL (1.18s vs. 1.98s). The difference is particularly notable for relational database sources, where PyRML outperformed RMLMapper on both MySQL (1.45s vs. 2.27s) and PostgreSQL (1.09s vs. 2.11s).
In addition to lower average execution times, PyRML also exhibited lower standard deviation across all source types, indicating more stable performance. For instance, in the case of CSV sources, the standard deviation for PyRML was 0.17s, compared to 0.25s for RMLMapper. This pattern is consistent for all data sources, reflecting the deterministic and efficient design of PyRML’s mapping engine.
These results highlight PyRML’s efficiency and robustness, confirming its suitability for interactive and automated data integration pipelines, especially where low-latency transformation is required.

\section{Conclusion and future work}
\label{sec:conclusion}
This paper introduced PyRML, a Python-native engine for declarative knowledge graph construction based on the RML specification. Our experimental evaluation demonstrates that PyRML is both standards-compliant and computationally efficient, achieving full or near-full conformance across a wide spectrum of data sources---including structured (CSV), semi-structured (JSON, XML), and relational databases (MySQL, PostgreSQL, SQL Server), as well as SPARQL endpoints. These results validate PyRML’s design, which closely mirrors the RML ontology and offers a transparent, modular, and extensible software architecture. When benchmarked against RMLMapper, one of the most widely adopted engines, PyRML exhibits consistently lower execution times and reduced performance variance, making it especially well-suited for automated or latency-sensitive knowledge graph construction pipelines. The native integration with Python’s data ecosystem (e.g., pandas, RDFLib, Jinja2) further enhances its usability for data scientists and knowledge engineers working in FAIR and open science contexts. 
Nevertheless, the evaluation highlights several im\-ple\-men\-ta\-tion-level limitations that are planned for resolution in upcoming releases. These include improved support for referencing object maps over SPARQL sources, quoting of SQL reserved keywords, handling of language-tagged literals, datatype conversions (e.g., xsd:boolean), and execution of custom SQL queries as logical sources. These are well-defined, bounded challenges that do not affect the core architecture, and their resolution is within reach in the short term.
Beyond these refinements, several promising directions emerge for future work. One major area of development is the extension of PyRML’s design and implementation to support RML extensions, such as those involving nested records, graph provenance, or incremental transformation. These features are increasingly relevant in contemporary use cases of knowledge graph construction, particularly in dynamic or federated data environments.
Another research trajectory lies in the integration of large language models (LLMs) into mapping workflows. As recent studies in knowledge engineering suggest~\cite{Lippolis2025}, LLMs can assist with the semi-automatic generation, validation, and explanation of mappings, potentially lowering the expertise required for building semantically rich data pipelines. We plan to explore how PyRML can serve as a backbone for such LLM-assisted workflows, offering human-in-the-loop interfaces and programmatic scaffolding for co-creating mappings that are both syntactically correct and semantically meaningful.
In conclusion, PyRML stands as a robust, efficient, and extensible platform for declarative RDF generation in Python. It addresses a key gap in the tooling ecosystem and opens up new possibilities for research and practice in transparent, scalable, and intelligent knowledge graph construction.

\begin{credits}
\subsubsection{\ackname} This work has been supported by the Water Health Open knoWledge (WHOW) project co-financed by the Connecting European
Facility programme of the European Union under grant agreement INEA/CEF/ICT/A2019/206322. Additional financial support to this project was provided by NextGenerationEU under NRRP Grant agreement n. MUR IR0000008 - FOSSR (CUP B83C22003950001).
\end{credits}
%
%
%
%
\bibliographystyle{splncs04}
\bibliography{references}
\end{document}